\documentstyle[12pt,twoside,fleqn,espcrc1,epsfig]{article}

\def\gtrsim{\ge}
\def\lesssim{\le}
\title{Strange particles in dense matter and kaon condensates}

\author{G.E. Brown, C.-H. Lee, and R. Rapp \address{Department of Physics, 
State University of New
York at Stony Brook, Stony Brook, NY 11794-3800, USA} 
\thanks{Supported by the U.S. Department of Energy under
grant no. DE-FG02-88ER40388.
We would like to thank Avraham Gal for useful criticism and advice.
One of us (RR) acknowledges support from the Alexander-von-Humboldt 
foundation as a Feodor-Lynen fellow.}  
}

\begin{document}

\maketitle

\begin{abstract}
We discuss the role of strangeness in dense matter and especially
in neutron stars. The early (in density) introduction of hyperons
found in many calculations is probably delayed by the decrease
in vector mean field acting on the neutron. This decrease results
from both conventional many-body rescattering effects and from the movement
towards asymptotic freedom at high densities.
Subthreshold $K^-$-meson production by the KaoS collaboration at GSI
shows that the $K^-$-mass must be substantially lowered, by $\gtrsim$
200 MeV at $\rho\sim 2\rho_0$.
It is shown that explicit chiral symmetry breaking through the kaon
mass may be responsible for $\Sigma^-$-nucleon and $\Xi^-$-nucleon 
scalar attraction being weaker than obtained by simple quark scaling.
The normal mode of the strangeness minus, charge $e^-$, excitation is
constructed as a linear combination of $K^-$-meson and $\Sigma^-$,
neutron-hole state. Except for zero momentum, where the terms are unmixed,
the ``kaesobar" is a linear combination of these two components.
\end{abstract}


\section{INTRODUCTION}

Strangeness has been introduced into dense matter, at supranuclear
densities, in many ways. Most simply, strange quarks have been included
in quark seas making up the neutron star. This seems energetically favorable,
since introducing strange quarks can relieve the high Fermi energies of
the nonstrange quarks \cite{bibref:1,bibref:2}. Bethe, Brown and Cooperstein
\cite{bibref:3} pointed out that such efforts gave a transition from hadrons
to quarks at much too low a density, because confining forces were left
out in the quark sector. Although we now believe that chiral restoration
in the up and down quarks takes place at $\rho\lesssim 3\rho_0$ \cite{bibref:4},
the strange quark condensate exists until substantially higher densities,
so it is more appropriate to use the hadron language for strange hadrons
at densities $\rho\sim 3\rho_0$ which we discuss here.

Breaking of strangeness takes only a fraction of a second, so that in
astrophysics there is sufficient of time for it except in the initial
bounce of a collapsing star, where dynamical times are $\sim 1$ millisecond.
Following the bounce the compact core of a large star goes through a period
of Kelvin-Helmholtz contraction, heating up as the trapped neutrinos in the
interior diffuse out of the core, leaving most of the energy in the core as 
heat. In the few seconds in which enough of the trapped neutrinos leave,
so that the phase transitions we discuss below can take place, the
temperature increases to $T\sim 50$ MeV.

The strange particles $K^-$, $\Sigma^-$ and $\Xi^-$ carry negative charge.
Thus, as the chemical potential $\mu_{e^-}$ goes up with increasing density,
it becomes favorable to replace electrons by strange particles, really
the negative charge being the important quantity.

\begin{figure}[htb]
\begin{minipage}[t]{70mm}
\epsfig{file=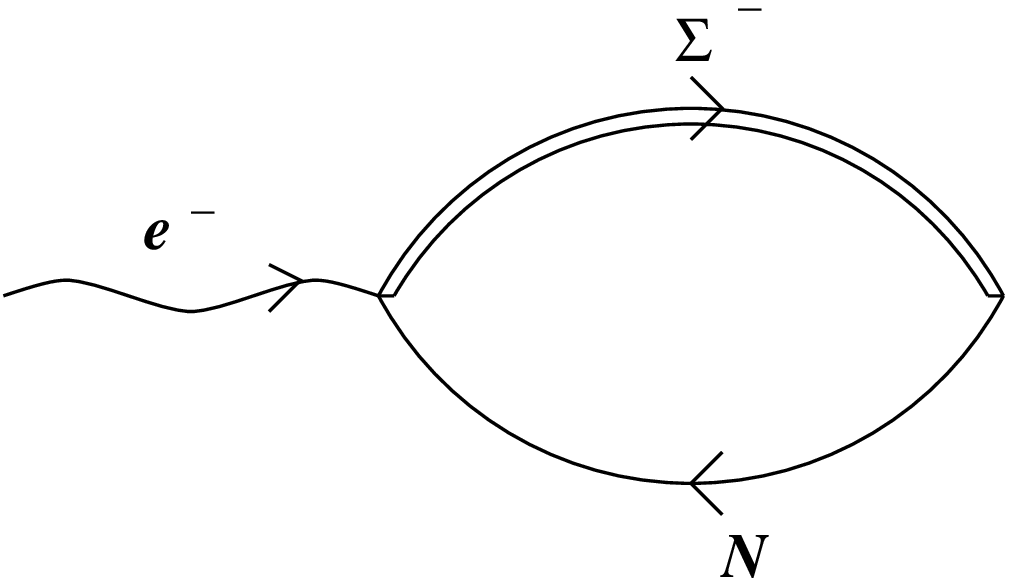,height=3.5cm}
\caption{The $\Sigma^- N^{-1}$ excitation can replace an electron
in a neutron star once the condition of Eq.~(\ref{eq:1.1}) is met.} 
\label{fig:1}
\end{minipage}
\hspace{\fill}  
\begin{minipage}[t]{85mm}
\epsfig{file=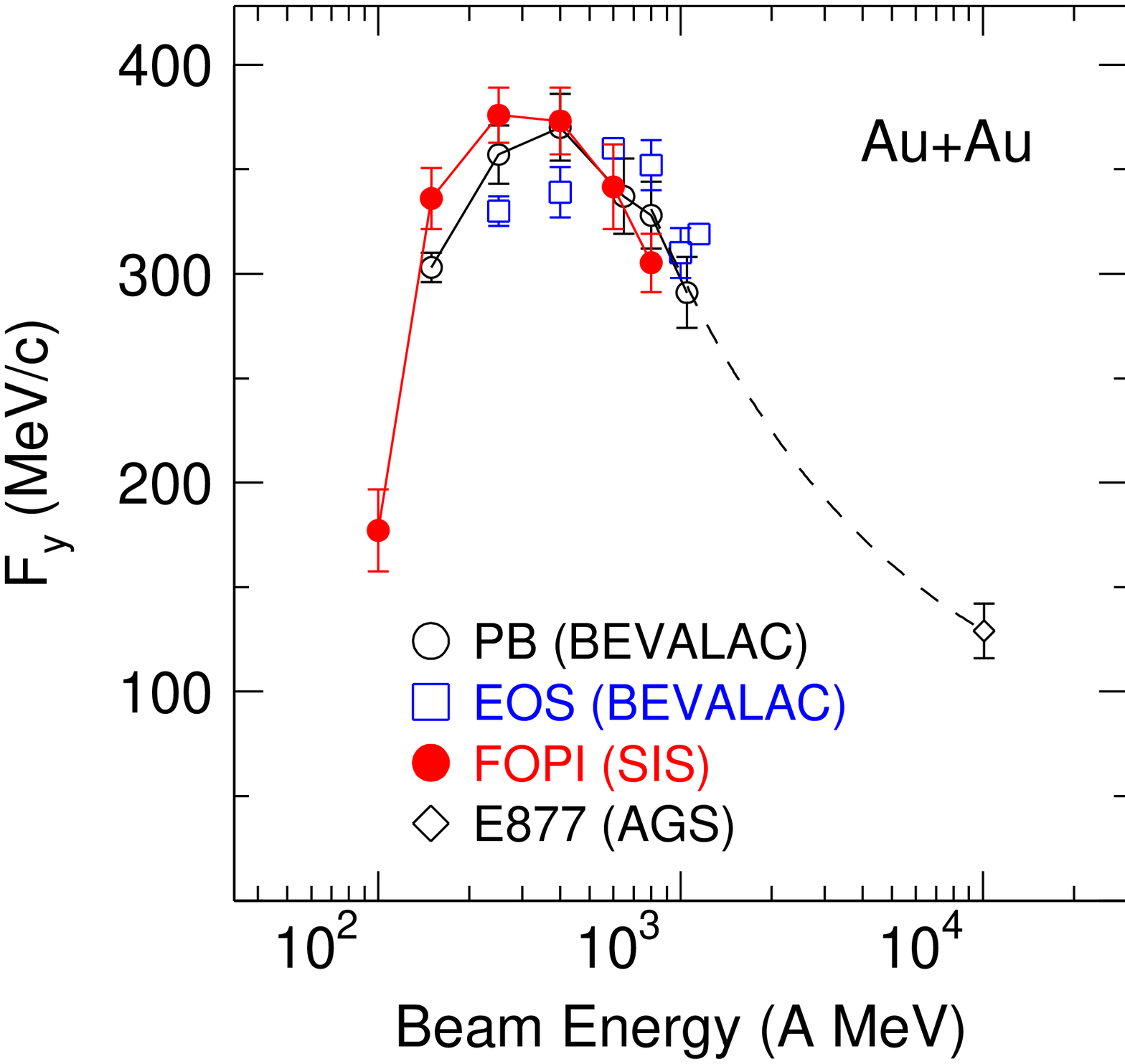,height=70mm}
\caption{The sideways flow as measured in Au$+$Au heavy ion collisions 
\protect\cite{bibref:10}.
}
\label{fig:2}
\end{minipage}
\end{figure}

Given linear, in density, extrapolation from nuclear matter density, it is
clear that $\Sigma^-$ baryons will quickly come in. In fact, $\Lambda$'s
generally enter first, but since they carry no charge, they effectively
change only little the behavior of the predominantly neutron star. The
$\Sigma^-$ can replace both neutron and electron. It comes in at the density
$\rho_c$ where
\begin{equation}
\mu_{\Sigma^-} =\mu_n +\mu_{e^-}.
\label{eq:1.1}
\end{equation}
In this case a $\Sigma^-$ at rest replaces a neutron at the top of its Fermi
sea, with Fermi momentum $k_{F_n}\sim 500$ MeV for $\rho\sim 3\rho_0$ and
an electron. It proves convenient to regard the $\Sigma^-$-particle,
neutron-hole as an excitation, as shown in Figure~\ref{fig:1}

Nowadays nobody makes the naive extrapolation in density we now outline,
but in a schematic way it represents the chief features of what goes on.
With $k_{F_n}\sim 500$ MeV (at $\rho\sim 3 \rho_0$) and
with chiral restoration at a density $\rho \le 3\rho_0$, all of the nucleon
mass here comes from the explicit chiral symmetry breaking through the low
quark masses \cite{LB98}
$m_n^\star\sim 200 - 300$ MeV, the neutron energy at the top of its
Fermi sea is
\begin{equation}
E_{F_n} \simeq \mu_n \gtrsim 500 \; {\rm MeV} + V
\end{equation}
where $V$ is the vector potential felt by the neutron. For $\rho\sim\rho_0$,
$V\simeq 275$ MeV in the linear Walecka model \cite{bibref:5}. Thus,
with a linear extrapolation to $\rho=3\rho_0$, $\mu_n\gtrsim 1325$ MeV.
{}From this it is clear that the condition Eq.~(\ref{eq:1.1}) is met at
a density $< 3\rho_0$, generally for $\rho\sim 2\rho_0$, even in more
complete theories.

Two points have been made clear by our model:
\begin{enumerate}
\item The $\Sigma^- N^{-1}$ excitation has a high momentum, $k\sim 500$ MeV
at $\rho\sim 3\rho_0$.
\item It's predominantly the vector mean field which pushes the neutron
energy up towards $M_{\Sigma^-}$ (and beyond in our schematic model).
\end{enumerate}

Now, two agencies operate so as to reduce the vector mean field at high
densities, one simple and well understood, the other somewhat tentative
and not yet well understood.

Firstly, rescattering corrections reduce a highly repulsive interaction
in many-body theory, essentially through screening. Thus, 
the Dirac-Brueckner-Hartree-Fock calculation \cite{bibref:6}, although 
consistent
with the Walecka model for $\rho=\rho_0$, reduces the vector mean field
by a factor $\sim 0.6$ for $\rho\sim 3\rho_0$. As can immediately be seen,
this moves the possible participation of $\Sigma^-$-hyperons from
$\rho\sim 2\rho_0$ to $\rho\sim 3 \rho_0$, and this is why we discuss
this latter density.

The second point has  to do with chiral restoration, basically asymptotic
freedom, and is not well understood. Thus, what we outline should be considered
as tentative.

Based on \cite{bibref:4} (which is certainly viewed in the community
as controversial), for $\rho=3\rho_0$ we are slightly beyond the 
critical density $\rho_c$ for chiral restoration. This means that in the
chiral limit (bare quark masses zero) the nucleon effective mass is zero.
Thus, nucleons behave as massless particles, interact via vector potentials.
In any case, nucleon flow in heavy ion collisions is determined chiefly
by the vector interactions. From the flow we can try to deduce the magnitude
of these interactions, in particular, from Au$+$Au collisions at 1 GeV/nucleon
carried out at GSI \cite{bibref:7}. 
Two groups \cite{bibref:8,bibref:9} have been
working to deduce the vector mean field from the data. This is not simple
to do, because the heavy-ion collisions are very nonequilibrium and large
explicitly momentum-dependent corrections must be made before one deduce
the vector mean field. Still, the two groups are in general agreement.
The more complete work of Sahu and Cassing \cite{bibref:9} places the vector
mean field $\sim 10\%$ below the DBHF one in magnitude at $\rho\sim 3\rho_0$,
namely $V=460$ MeV at this density. That of \cite{bibref:8} places it
slightly above the DBHF value. Note that this DBHF value is arrived at
by standard rescattering corrections to the mean field extrapolated linearly
from $\rho=\rho_0$.

Empirically, as can be seen in Figure~\ref{fig:2}, the flow in Au$+$Au 
collisions is already decreasing by the time $E_{beam}/A=1$ GeV.
At this energy the transport calculation show that the density 
$\rho\sim 3\rho_0$ is reached. Higher densities are reached at higher
bombarding energies.

{}From the figure it is clearly seen that the flow decreases rather
strongly above $E=1$ GeV/N. Sahu and Cassing model this decrease by a
vector mean field which goes to zero at higher densities.

Since, in the transport calculation \cite{bibref:8,bibref:9}, densities of
$\rho\sim 3\rho_0$ are reached for Au$+$Au at 1 GeV/N and since, according
to \cite{bibref:4}, chiral restoration has already 
occurred by this density,
we interpret the decrease in vector mean field with increasing density
as a manifestation of asymptotic freedom. Indeed, from lattice calculations
of the quark number susceptibility, at zero baryon chemical potential, it is
 seen that the effective vector coupling decreases by an order of magnitude
-- but does not go to zero -- in going 
through the phase transition \cite{bibref:11}.
In this reference it is made clear that the (Fierzed) colored gluon exchange
takes over from the $\omega$-meson exchange as the temperature goes up
through the phase transition. Thus, we have an effective vector mean field 
remaining above $T_c$. One would expect the same general behavior with density,
except that the decrease in vector mean field does not seem to be so rapid
with density as with temperature.

Given the behavior of the vector mean field described above, we feel that
the most favorable density for hyperons in nuclei is $\rho\sim 3\rho_0$.
As asymptotic freedom sets in at higher densities, the situation will become
less favorable.

\section{NUCLEON-HYPERON INTERACTION}

In discussing interactions between the nucleon and hyperons, those from the
exchange of vector particles should be straightforward. Beginning with
the vector dominance of Gell-Mann, Sharp and Wagner \cite{bibref:12} 
and continuing
through the theories in which vector mesons are viewed as gauge particles of
hidden symmetries \cite{bibref:13}, the $\omega$-meson is coupled to the baryon
number in nonstrange quarks, and the $\rho$-meson by SU(3) symmetry, simply
related to the $\omega$-coupling ($9 g_{\rho NN}^2= g_{\omega NN}^2$).

\begin{figure}[htb]
\begin{minipage}[t]{75mm}
\epsfig{file=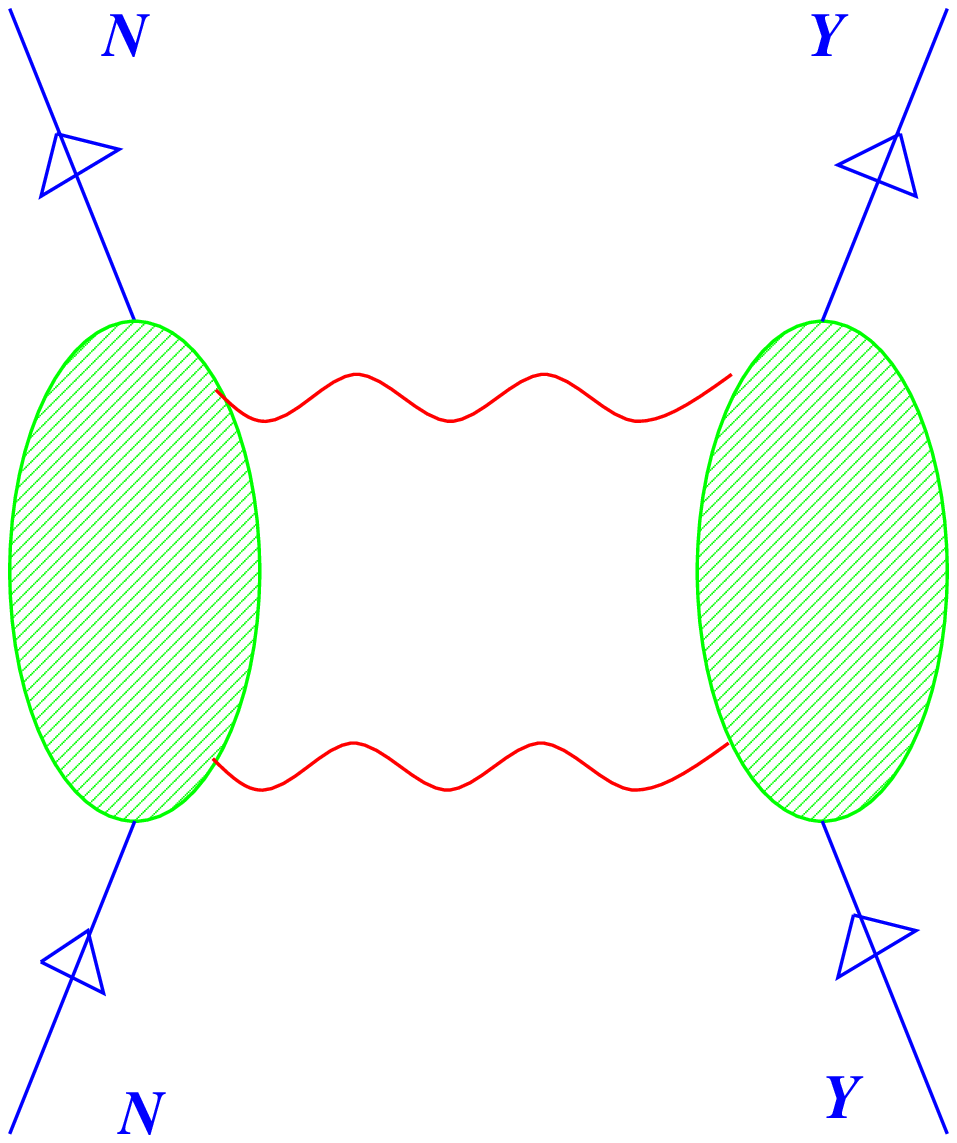,height=7cm}
\caption{Correlated $\pi\pi$ or $K\bar K$ interactions, the wavy lines showing
the exchanged mesons. The shaded blobs stand for all possible intermediate
states.}
\label{fig:3}
\end{minipage}
\hspace{\fill}  
\begin{minipage}[t]{80mm}
\epsfig{file=optpot.epsi, height=10cm}
\caption{$\Sigma^-$ optical potentials of \protect\cite{bibref:17}
in symmetric nuclear matter. Here we used the reduced $\Sigma^-$-nucleon mass 
as free $\Sigma^-$ mass as mentioned in \protect\cite{bibref:21}.
For neutron dominant matter, $\Sigma^-$ feels more repulsion. 
The upper curves from the density-dependent analyses are favored.
The lower curves are for \underline{no} density dependence, i.e. an
optical potential linear in the density.}
\label{fig:4}
\end{minipage}
\end{figure}

In the case of the scalar $\sigma$-meson, the situation is more complicated.
The scalar attraction which binds nuclei results from an enhancement, through
scalar interactions, of two-pion exchange, the two pions being correlated
in a relative S-state. For nucleon-hyperon interactions, correlated $K\bar K$
states also enter, becoming relatively more important as the strangeness
of the hyperon increases.
The correlated $\pi\pi$ or $K\bar K$ interactions are shown in 
Figure~\ref{fig:3}.

The scalar interactions have been calculated for many years now in the case 
of the nucleon-nucleon potential, and are well understood. Recently, the
calculation of the nucleon-hyperon interaction by boson exchange has been
taken up by the J\"ulich group \cite{bibref:14}.
Coupling constants of the $\pi$ and $K$ to nucleons and hyperons are determined
by SU(3) symmetry. Energy denominators etc. are taken empirically and involve 
a lot of SU(3) symmetry breaking, arising chiefly from the relatively large
strange quark mass $m_s\sim 150$ MeV, which results in the large kaon mass
$m_K\simeq 495$ MeV. The large kaon mass means that kaon exchange is somewhat
suppressed with respect to pion exchange, so there is no simple rule, as in
the case of vector meson exchange, for obtaining the couplings of the effective
$\sigma$ by quark counting. None the less, there is an effective symmetry,
which we can discover by removing the explicit symmetry breaking in the meson
propagators, the wavy lines in Figure~\ref{fig:3}.

Let us consider how the kaon mass diminishes the kaon exchange in processes 
such as shown in Figure~\ref{fig:3}. The kaon propagator is
\begin{equation}
D_K (q) =\frac{1}{q^2+m_K^2-\omega^2}.
\label{eq:2.1}
\end{equation}
We neglect recoil, setting $\omega=0$. (Since off-diagonal matrix elements
like $\Lambda+\pi \longrightarrow \Sigma$ in intermediate energy states
are involved, this may be an even cruder approximation than in the case of the
nucleon-nucleon interaction.) Now $|\vec q|$ will be limited by the form
factor, which generally has a scale of $\Lambda\sim 1$ GeV $\sim 2 m_K$.
Thus, the $m_K^2$ in the denominator of Eq.~(\ref{eq:2.1}) should be
an $\sim$ 25\% effect. Given that two meson exchanges are involved in the
interaction, the explicit chiral symmetry breaking through the presence of
the kaon mass in the denominators roughly estimated to be an $\sim 50\%$
effect. Of course there is further chiral symmetry breaking in the ``blob"
in Figure~\ref{fig:3}, which we neglect.

\begin{table}[hbt]
\newlength{\digitwidth} \settowidth{\digitwidth}{\rm 0}
\catcode`?=\active \def?{\kern\digitwidth}
\caption{Effective $\sigma$ coupling strength for correlated $\pi\pi$ and
$K\bar K$ exchange for various baryon-baryon channels, from 
\protect\cite{bibref:14}.}
\label{tab:1}
\begin{tabular*}{\textwidth}{@{}l@{\extracolsep{\fill}}cccc}
\hline
 & $NN$ & $N\Lambda$ & $N\Sigma$ & $N\Xi$ \\
\hline
Full Model  & 5.87 & 2.82 & 2.58 & 1.19 \\
$(K\bar K)$ & 1.00 & 1.02 & 1.52 & 0.84 \\
1.5 $(K\bar K)$ & 1.50 & 1.53 & 2.28 & 1.26 \\
Full Model$\star$ & 6.52 & 3.33 & 3.34 & 1.61\\
\hline
\multicolumn{5}{@{}p{120mm}}{
$\star$ Full model with $(K\bar K)$ replaced by $1.5 (K\bar K)$.}
\end{tabular*}
\end{table}

The most extensive meson exchange analyses have been performed by
Reuber\cite{bibref:14}. We show these in Table~\ref{tab:1}, where the full
model calculations are shown, also those from correlated $K\bar K$ exchange.
Note that the latter become more important with increasing strangeness, a fact
that will turn out to be most important.

In the last row here we have tried to reconstruct what the interactions would
be were the kaon mass equal to zero.
In fact, with our coupling of the vector interactions to the baryon number
in nonstrange quarks, the $N\Lambda$ coupling shown in Table~\ref{tab:1}
is unsatisfactory. In order to reproduce the known $\Lambda$ binding
energy in heavy hypernuclei, Ma et al. \cite{bibref:15} had to use a
ratio of $g_{\Lambda \Lambda\omega}/g_{NN\omega}$ of $0.512$ instead
of the $2/3$ we propose. If the $2/3$ is used, then the $N\Lambda$
coupling must be increased to $\gtrsim 0.6$ (In \cite{bibref:16}
the $N\Lambda$ coupling is 0.619.) We do not know the underlying origin of 
this, but the evidence of this larger $N\Lambda$ coupling is persuasive.
(The $\Lambda$ binding energies only determine the ratio
of $g_{\Lambda\Lambda\omega}/g_{\Lambda\Lambda\sigma}$ \cite{GM91}.
Thus, phenomenology does not give a clue to whether $g_{\Lambda\Lambda\sigma}$
is large or small, unless $g_{\Lambda\Lambda\omega}$ is fixed by theory, or
vice versa.)

Our theme is that the explicit symmetry breaking by the kaon mass decreases the
$N\Sigma$ coupling more than the $N\Lambda$ one, and proportionately decreases 
the $N\Xi$ coupling, so that roughly
\begin{equation}
\frac{g_{N\Sigma}}{g_{N\Lambda}} \sim 0.9 ; \;\;\;
\frac{g_{N\Xi}}{g_{N\Sigma}/2} \sim 0.9 ,
\label{eq:2.2}
\end{equation}
certainly valid to within the accuracy of the calculations. (Remember
that the interactions go with the square of the coupling constant.) 
In Figure~\ref{fig:4}
we show the $\Sigma^-$-potential which would be obtained from the parameters 
in Table~11 of \cite{bibref:17}. The repulsive potentials result from the
density-dependent optical model potentials which the authors of 
\cite{bibref:17} favor. The density dependent potentials give a strong
repulsion at $\rho\sim 3\rho_0$. Extrapolation to these high densities
may not be warranted. In any case, the repulsions sum to rule out 
$\Sigma^-$-hypernuclei \cite{bibref:17}. We suggest, on the basis of our
reconstruction of the interaction in the chiral limit that at least some of
the difference between $\Lambda$ and $\Sigma^-$ couplings derives from the
explicit chiral symmetry breaking.

Emulsion data, summarized in \cite{bibref:20}, showed a consistency for a
$\Xi$-nucleus potential depth of $\sim 24$ MeV attraction. This has recently
been challenged by KEK quasifree production of the $\Xi$ in $^{12}$C where it
is claimed that the potential, still attractive, is perhaps shallower, down
to $\sim 15$ MeV. In Balberg and Gal \cite{bibref:21}, the $\sim 15$ MeV was
assumed. With this assumption the cascade particle replaced the $\Sigma^-$
rather rapidly as a function of density.

The results in Table~\ref{tab:1} disfavor the $\Xi^-$ as compared with
the $\Sigma^-$ in that the attraction is smaller, cut down by a factor
$\sim (0.9)^2$ (See Eq.~(\ref{eq:2.2})). The role played by the $\Xi^-$
is uncertain, since its interaction with nucleons is not pinned down
well empirically.

\section{KAON INTERACTION IN DENSE MATTER AND KAON CONDENSATION}

The attraction resulting from the partial restoration of explicit
chiral symmetry breaking felt by kaons in nuclear matter was a surprise
when first presented by Kaplan and Nelson \cite{bibref:22}. It was recently
argued by Brown and Rho \cite{bibref:23}
that at supranuclear densities, this attractive scalar field could
be considered to be the same sort of scalar field we considered in
section 2 between nuclear matter and the nonstrange antiquark in the kaon.
The difference from hyperonic interaction in dense matter is that for the
$K^-$-meson, composed of up-antiquark and strange quark, the vector
interaction, because of G-parity, gives an attractive interaction on the
nonstrange antiquark. Thus, the scalar and vector interactions add in
magnitude, both furnishing attraction. Clearly this attraction is expected
to be large in magnitude.

Much has been written about $K^-$-nuclear matter interactions, and has
most recently been summarized in \cite{bibref:24}. Important experimental
results have recently been published by the KaoS collaboration. In
Ni$+$Ni at 1.8 GeV, densities up to $\sim 2\rho_0$ are reached.
(The precise density depends, of course, on the mean field used in the
transport calculations, so even this is somewhat model dependent.) Results for
subthreshold $K^-$-production are shown in Figure~\ref{fig:5}. The mean fields
used to calculate in-medium effects are shown in Figure~\ref{fig:6}.

\begin{figure}[htb]
\begin{minipage}[t]{77mm}
\epsfig{file=akni18_color.epsi,height=90mm}
\caption{$K^-$ spectra in Ni$+$Ni collisions at 1.8 A GeV beam kinetic
energy.}
\label{fig:5}
\end{minipage}
%
\hspace{\fill}
%
\begin{minipage}[t]{78mm}
\epsfig{file=kmass_color.epsi,height=9cm}
\caption{Effective mass of kaon and antikaon in nuclear medium.}
\label{fig:6}
\end{minipage}
\end{figure}

There is general agreement between the theoretical groups of \cite{bibref:24}
and that of Cassing et al. \cite{bibref:27} on the fits to the $K^-$-spectra.
In a footnote in \cite{bibref:27} the $K^-$-mass drops slightly more rapidly
than in our Figure~\ref{fig:5} (although uncertainties are emphasized there).

The KaoS experiments are particularly useful in investigation of
mean field effects. The $K^-$-production would be 235 MeV below threshold
for 1.8 A GeV Ni$+$Ni, whereas a mean field such as produces
$m_{\bar K}^\star$ shown in our Figure~\ref{fig:5} would put the production
locally at threshold at the $\sim 2\rho_0$ density formed in central collisions.
The extrapolation of the curve from $2\rho_0$ to $3\rho_0$ is not determined by
the present experiments, but will be checked by subthreshold $K^-$-production
in Au$+$Au collisions, which do reach $\rho\sim 3\rho_0$ and which are now 
going on.  Note that the subthreshold kaon production is insensitive to the
low densities involved in $K^-$-atoms and do not bear on the results of
Friedman, Gal and Batty \cite{bibref:28}.

A dramatic result of medium effects is shown in Figure~\ref{fig:7}, where
the ratio of $K^+/K^-$ production, each chosen at an energy
which would be 235 MeV below threshold in nucleon-nucleon collisions.
It is seen that this ratio is changed by an order of magnitude by medium
effects. This is because the $K^+$-meson experiences scalar and vector
mean fields of opposite sign, which tend to cancel each other.

\begin{figure}[htb]
\begin{minipage}[t]{77mm}
\epsfig{file=kakr_color.epsi,height=90mm}
\caption{Kinetic energy spectra of $K^+/K^-$
in Ni+Ni collisions at 1.0 AGeV ($K^+$) and
1.8 AGeV ($K^-$). The solid and dotted histograms are the
results with and without kaon medium effects,
respectively. The circles are the experimental data
from the KaoS collaboration \protect\cite{bibref:25}.
}
\label{fig:7}
\end{minipage}
%
\hspace{\fill}
%
\begin{minipage}[t]{78mm}
\epsfig{file=kasob.epsi,height=89mm}
\caption{Lowest branches of kaesobar at $\rho=3\rho_0$. 
Lines I and II represent the lowest
$\Sigma^-$-particle-neutron-hole branches corresponding to
$V_{\Sigma^-}(3\rho_0)=250$ MeV, 0 MeV, respectively.
The solid lines are the possible lowest kaesobars which 
follow the minimum energy solution between $K^-$ and 
$\Sigma^-$-particle-neutron-hole branch.
}
\label{fig:8}
\end{minipage}
\end{figure}

Now the electron chemical potential $\mu_e$ is $\sim 210$ MeV at
$\rho\sim 3\rho_0$ \cite{bibref:29}. As soon as the $K^-$-mass drops down
to $\mu_e$, kaons will replace electrons in neutron stars, with many 
dramatic consequences, as outlined in \cite{bibref:29,bibref:30}.

There have been many articles in the literature, most recently 
Ref.~\cite{bibref:31}, saying that the introduction of hyperons early in density
precludes kaon condensation by lowering the electron chemical potential.
Whereas it is clear that the two schemes are somewhat competitive, one
should realize that at the temperature involved in stars, $T\sim 50$ MeV,
all normal modes with electron charge will be excited with appropriate
Boltzmann factors. The $K^-$ and $\Sigma^-$--neutron hole
($\Sigma^- N^{-1}$) interact, except at momentum $k=0$, and one should
diagonalize these excitations in order to find the normal mode, the
kaesobar \cite{bibref:24}. We show this construction in Figure~\ref{fig:8}.
Lines I and II represent the lowest possible 
$\Sigma^-$-particle-neutron-hole branches corresponding to
$V_{\Sigma^-}(3\rho_0)=250$ MeV, 0 MeV, respectively.
The solid lines are the lowest possible kaesobars which 
follow the minimum energy solution between $K^-$ and the  
$\Sigma^-$-particle-neutron-hole branch.
Here we used $\Phi(3\rho_0) = 700$ MeV, $W(3\rho_0) = 460$ MeV, 
$R(3\rho_0)=-51$ MeV,
$x_{\sigma K} = x_{\omega K} = 1/3$, $x_{\rho K} = 1/9$,
and $x_{\sigma\Sigma}=0.5$ in the meson exchange model (see Appendix
of \protect\cite{bibref:24}). 
In order to specify the  $\Sigma^-$ potential, we 
used
$x_{\omega \Sigma}=x_{\rho\Sigma}=1.174$, 0.667 for lines I and II,
respectively. 
(For case I the full mean field potential felt by the $\Sigma^-$,
vector plus scalar, is 250 MeV. Given the mean fields $\Phi$, $W$,
$R$ these $x_{\omega\Sigma}$ and $x_{\rho\Sigma}$ are what 
we needed to reproduce the two curves.)
The value $V_{\Sigma^-}(3\rho_0)= 250$ MeV for curve I is taken from the
least repulsive of the upper curves of Figure~\ref{fig:4}. Curve II lies
much lower, since we used 2/3 of the reduced vector field of
$W=460$ MeV on the nucleon but 1/2 of the very large (attractive)
700 MeV scalar potential on the nucleon which is typical of
Walecka mean field theories. Although curve II lies much lower than curve I,
we believe that such a scenario might result from theories which have
the strong scalar attraction of Walecka-like theories.

In the sense that the upper curve in Figure~\ref{fig:8} follows from the
least repulsive mean field in the extrapolation of empirical data 
\cite{bibref:17}, one might think that the $\Sigma^-$ does not enter
in dense matter. However, scalar mean fields of the size we employed
in case II are needed for chiral restoration at $\rho \lesssim 3\rho_0$.
Furthermore we have argued that the vector mean fields are substantially
reduced from those of the linear Walecka model.
We offer case II as a model which roughly incorporates this features.


\end{document}